\documentclass[a4paper]{PoS}

\def\<{\left\langle}
\def\>{\right\rangle}

\newcommand{\tr}{ {\rm Tr} \, }
\newcommand{\Tr}[1]{ {\rm Tr}\left( #1 \right) }

\def\opsi{\overline{\psi}}

\def\hmu{\hat{\mu}}
\def\hnu{\hat{\nu}}
\def\muB{\mu_q}
\def\dmu{\partial (\mu_q a)}

\title{Lattice simulation of $QC_2D$ with $N_f=2$ at non-zero baryon density}

\ShortTitle{Lattice simulation of $QC_2D$ with $N_f=2$ at non-zero baryon density}

\author{V.V.~Braguta,$^{abc}$ A.Yu.~Kotov,$^c$ \speaker{A.A.~Nikolaev},$^b$ and S.N.~Valgushev$^c$\\
        \llap{$^a$}Institute for High Energy Physics, National Research Centre "Kurchatov Institute"\\
        pl. Nauki 1, Protvino, Moscow region, 142281 Russia\\
        \llap{$^b$}School of Biomedicine, Far Eastern Federal University\\
        Sukhanova 8, Vladivostok, 690950 Russia\\
        \llap{$^c$}Institute for Theoretical and Experimental Physics\\
		Bolshaya Cheremushkinskaya 25, Moscow, 117218 Russia\\
        E-mail: \email{braguta@itep.ru}, \email{kotov@itep.ru}, \email{nikolaev.aa@dvfu.ru}, \email{valgushev@itep.ru}}

\abstract{The lattice simulations of $QC_2D$ with two flavors of staggered fermions and non-zero quark chemical potential $\mu_q$ have been performed. Dependencies of the Polyakov loop, chiral condensate and baryon number density on $\mu_q$ were studied. We found that an increase of the baryon chemical potential leads to chiral symmetry restoration. At small values of $\mu_q$, our results for the baryon number density agree with ChPT predictions.}

\FullConference{The 33rd International Symposium on Lattice Field Theory\\
		14 -18 July 2015\\
		Kobe International Conference Center, Kobe, Japan*}

\begin{document}

\section{Introduction}
The study of the phase diagram of QCD is very important for astrophysics, cosmology, and particle physics. However, the phase diagram of QCD is still incompletely studied. The reason is that quarks and gluons constitute a strongly interacting system, and this fact complicates the use of analytical methods. One of the possible approaches to the study of the properties of such a system is the LQCD, where functional integrals are numerically calculated using the Monte Carlo method. Calculations in lattice QCD allow ab initio study of the properties of a quark-gluon plasma~\cite{bib:A_Bazavov}.

However, calculations in LQCD with finite chemical potential are now impossible due to the sign problem of the fermionic determinant. Therefore, we shall study a simpler theory: QFT with $SU(2)$ gauge group and two degenerate flavors, where no sign problem arises~\cite{bib:JB_Kogut_sign,bib:T_Makiyama,bib:S_Hands_sign}, in order to better understand the qualitative features of the QCD phase diagram and to analyze the effect of a non-zero chemical potential on the properties of QGP. The point is that $QC_2D$ has specific relation for the Dirac operator~\cite{bib:JB_Kogut_sign,bib:T_Makiyama}:
\begin{eqnarray}\label{Dirac_relation}
det\Bigl[ M(\muB) \Bigr] = det\Bigl[ \left( \tau_2 C \gamma_5 \right)^{-1} M(\muB) \left( \tau_2 C \gamma_5 \right) \Bigr] = det\Bigl[M(\muB^*)\Bigr]^*,
\end{eqnarray}
where $M(\muB) = \gamma_\mu D_\mu + m_q - \muB \gamma_4$ is the Dirac operator in continuum $R^4$, $\mu_q = \mu_B / 2$ is the quark chemical potential, $C = \gamma_2 \gamma_4$, and $\tau_2$ is a generator of the $SU(2)$ group. Relation~\ref{Dirac_relation} guarantees, that $det\Bigl[ M(\muB) \Bigr]$ is real for the real $\muB$. One can also prove~\cite{bib:S_Hands_sign}, that the spectrum of $M(\muB)^\dagger M(\muB)$ at finite real $\muB$ is strictly positive, both in the continuum and for the lattice formulation~\ref{Dirac_operator}, if the quark mass is non-zero.

Our aim is to study the influence of the chemical potential on the Polyakov loop, chiral condensate and baryon number density. Especially interesting is to understand the effect of a non-zero baryon density on the breaking/recovery of the chiral symmetry. Similar investigations were performed in~\cite{bib:T_Makiyama,bib:S_Cotter,bib:JI_Skullerud} for $N_f = 2$ with Wilson fermions and in~\cite{bib:JB_Kogut_1}\cite{bib:JB_Kogut_2} with $N_f = 4$ and 8 flavors of staggered fermions respectively. However, Wilson fermions explicitly violate the chiral symmetry~\cite{bib:Gattringer_Lang}, thus they may not reveal all the phase transition lines in the $QC_2D$ phase diagram. In this paper we consider $N_f=2$ flavors of staggered fermions, taking the fourth root of the $det\Bigl[ M(\muB)^\dagger M(\muB) \Bigr]$ via the R-algorithm~\cite{bib:HJ_Rothe}.

\section{Lattice formulation}
The partition function of the system under study has the form:
\begin{eqnarray}\label{Z}
Z = \int DU\,det\Bigl[M^\dagger(\muB) M(\muB) \Bigr]^{ \frac{1}{4} } e^{- S_G[U]},
\end{eqnarray}
where the functional integration is performed over the $SU(2)$ group manifold, $M(\muB)$ is the lattice Dirac operator for
Kogut-Susskind fermions with the baryon chemical potential, and $S_G[U]$ is the Wilson gauge action~\cite{bib:KG_Wilson}:
\begin{eqnarray}\label{S_G}
S_G = \beta \sum_{x}\sum_{\mu < \nu = 1}^4 \Bigl(1 - \frac{1}{2} \tr U_{x, \mu\nu} \Bigr).
\end{eqnarray}
Here $\beta = \frac{4}{g^2}$, and $U_{x, \mu\nu} = U_{x, \mu} U_{x + \hmu, \nu} U^\dagger_{x + \hnu, \mu} U^\dagger_{x, \nu}$. The lattice Dirac operator $M(\muB)$ in~\ref{Z} has the form:
\begin{eqnarray}\label{Dirac_operator}
M_{xy} = ma\delta_{xy} + \frac{1}{2}\sum_{\mu = 1}^4 \eta_\mu(x)\Bigl[ U_{x, \mu}\delta_{x + \hmu, y}e^{\muB a\delta_{\mu, 4}} - U^\dagger_{x - \hmu, \mu}\delta_{x - \hmu, y}e^{- \muB a\delta_{\mu, 4}} \Bigr],
\end{eqnarray}
where $a$ is the lattice spacing, $m$ is the mass of the quark, and the functions $\eta_1(x) = 1,\, \eta_2(x) = (-1)^{x_1},\, \eta_3(x) = (-1)^{x_1 + x_2},\, \eta_4(x) = (-1)^{x_1 + x_2 + x_3}$ are the $\gamma$-matrices after the Kogut-Susskind transformation.
The chemical potential $\muB$ is introduced in~\ref{Dirac_operator} by means of the multiplication of time links by the exponential factor $e^{\pm \muB a}$. This way of the introduction of the chemical potential makes it possible to avoid additional divergences and to reproduce the known result for free fermions in the continuum limit~\cite{bib:RV_Gavai}.

\begin{figure}[t]
\begin{center}
	\begin{minipage}[t]{0.49\textwidth}
    		\includegraphics[width = 1.0\textwidth]{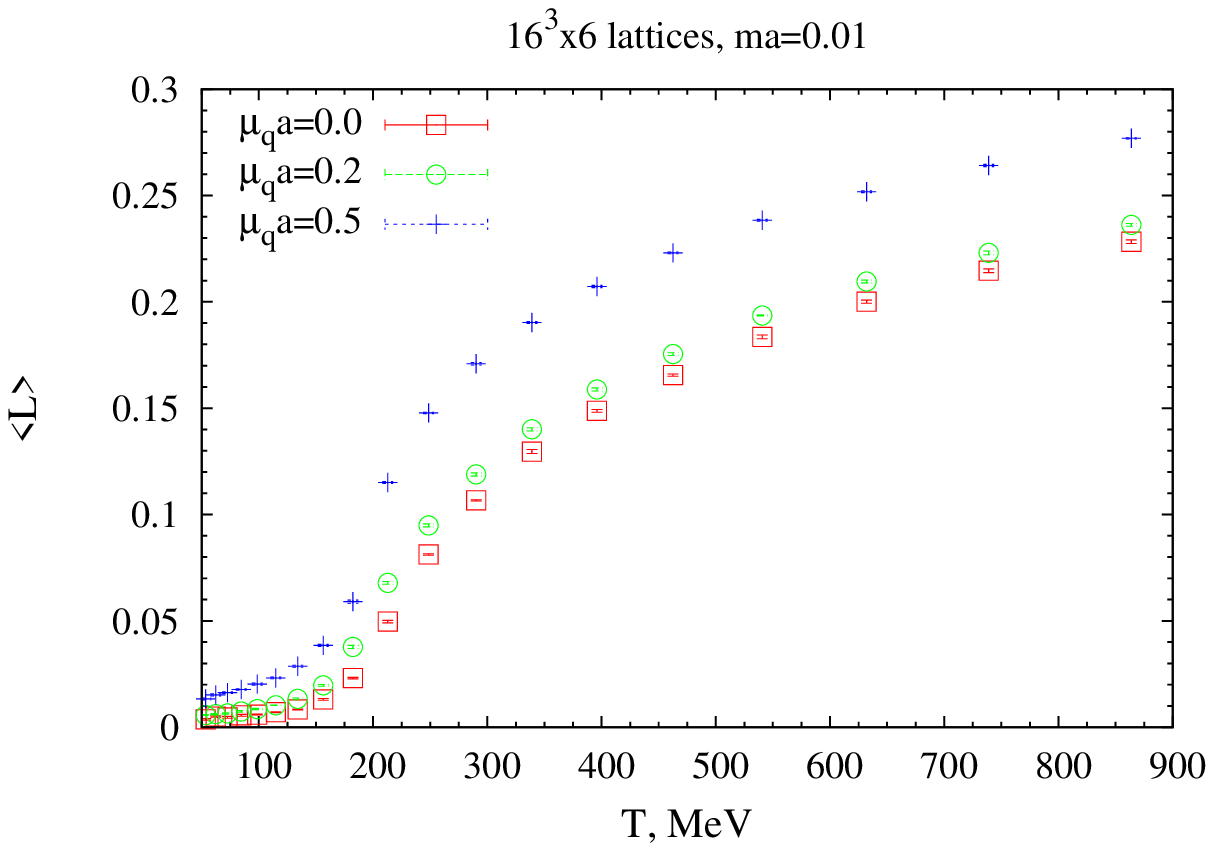}
\caption{Polyakov loop as a function of T for three values of the baryon chemical potential.}
 		\label{fig:P_loop}
    \end{minipage}
\hfill
    \begin{minipage}[t]{0.49\textwidth}
    		\includegraphics[width = 1.0\textwidth]{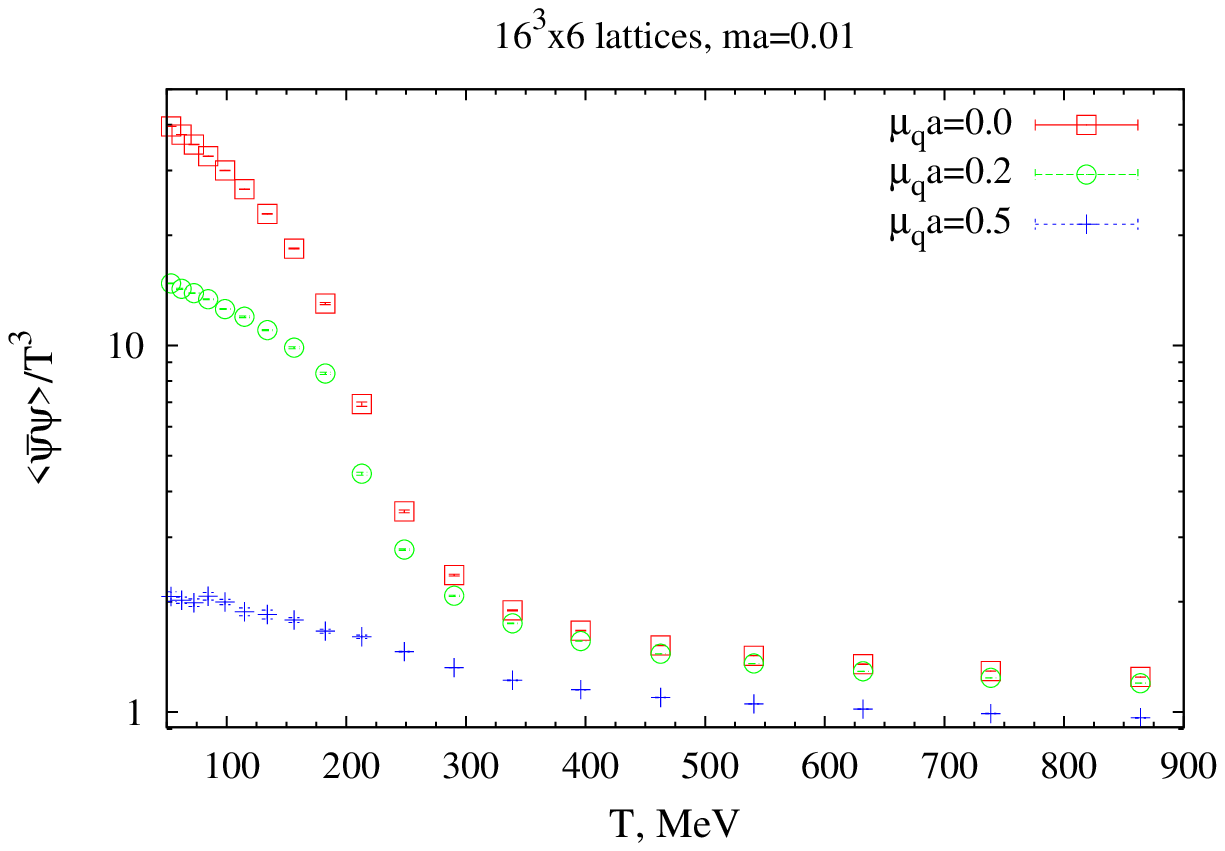}
\caption{Chiral condensate as a function of T for three values of the baryon chemical potential. The ordinate axis
is given on a logarithmic scale.}
 		\label{fig:Ch_cond}
    \end{minipage}
\end{center}
\end{figure}

For partition function~\ref{Z} in the continuum limit to correspond to two dynamic quark flavors, we extract the fourth order root of the fermionic determinant using the rational approximation with an accuracy of $O(10^{-15})$~\cite{bib:MA_Clark}. Configurations were generated by means of the hybrid Monte Carlo method, $\Phi$-algorithm~\cite{bib:Gattringer_Lang} was employed. We considered a $16^3 \times 6$ lattice with the bare fermion mass $ma = 0.01$, $\beta$ = 1.6 \ldots 2.7, and $\muB a$ = 0.0 \ldots 0.6 (for each set of parameters, 400 independent configurations were generated). The program code was written with the use of CUDA. The calculations were performed at the ITEP supercomputer and IHEP cluster.

\begin{figure}[t]
\begin{center}
    	\includegraphics[width = 0.75\textwidth]{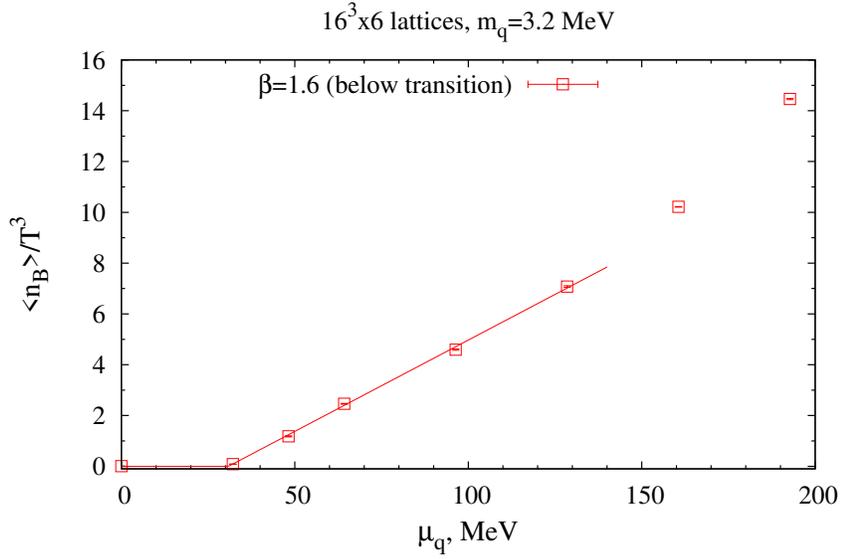}
	\caption{Baryon number density as a function of~$\muB$ in the low-temperature phase ($\beta = 1.6$). Linear fit is shown.}
 	\label{fig:Bar_density}
\end{center}
\end{figure}

\section{Numerical results and discussion}
To study the physical properties of the system, we considered the following observables (triangular brackets mean thermodynamic averaging):
\begin{itemize}
	\item Polyakov loop:
\begin{eqnarray}\label{Pol_loop}
\< L \> = \frac{1}{N_s^3} \sum_{x_1,x_2,x_3 = 0}^{N_s - 1} \frac{1}{2} \< \tr \prod_{x_4 = 0}^{N_\tau - 1} U_{x, 4} \> ;
\end{eqnarray}	
	\item chiral condensate:
\begin{eqnarray}\label{chiral_condensate}
a^3\<\opsi\psi\> = - \frac{1}{N_s^3 N_\tau}\frac{1}{8}\frac{\partial}{\partial (ma)} log\,Z = \frac{1}{N_s^3 N_\tau}\frac{1}{8} \< \tr M^{-1} + \tr (M^\dagger)^{-1} \> ;
\end{eqnarray}
	\item baryon number density:
\begin{eqnarray}\label{bar_number}
a^3 \<n_B\> = \frac{1}{N^3_s N_\tau} \frac{1}{16} \frac{\partial (log\,Z)}{\dmu} = \frac{1}{N^3_s N_\tau} \frac{1}{8} \< Re \Tr {\frac{\partial M}{\dmu} M^{-1}} \> .
\end{eqnarray}	
\end{itemize}

In order to fix the scale we employed Sommer parameter $r_0$ = 0.468(4) fm~\cite{bib:A_Bazavov} and performed the measurements on $16^3 \times 32$ lattices with $m_q a = 0.01$. Lattice spacings and pion masses are listed in the Table~\ref{tabular:pion_masses}.
\begin{table}[bH!]
\begin{center}
\begin{tabular}{|c|c|c|}
\hline
$\beta$ & $a, fm$ & $M_{\pi}, MeV$ \\ 
\hline
1.7 		& 0.45(1)  & 109(3)  \\
1.9 		& 0.20(1)  & 216(6)  \\
2.1 		& 0.135(2) & 430(13) \\
2.2 		& 0.097(1) & 551(16) \\
\hline
\end{tabular}
\end{center}
\caption{Lattice spacings and pion masses.}
\label{tabular:pion_masses}
\end{table}

Figure~\ref{fig:P_loop} shows the dependences of the Polyakov loop on temperature for three $\muB$ values, a crossover phase transition is observed. It may also be seen in the figure, that an increase in the baryon chemical potential results in an slight increase in $\< L \>$ for the same temperature. However, the susceptibilities of $\< L \>$ can not be measured with the existing statistics and the influence of the baryon chemical potential on $T_c$s cannot be determined.

Figure~\ref{fig:Ch_cond} shows the dependences of the chiral condensate on $T$ for various $\muB$ values. It can be seen that an increase in the baryon chemical potential leads to a significant decrease in $\< \overline{\psi} \psi \>$, i.e., to the recovery of the chiral symmetry. These results are in agreement with the previous results obtained for Wilson fermions~\cite{bib:S_Cotter} and with the outcomes of~\cite{bib:JB_Kogut_ChPT}. On the figure~\ref{fig:Bar_density} the dependence of the baryon number density on $\muB$ in the confinement phase is shown. One can see, that at $\muB \approx m_\pi / 2$ it begins to rise linearly, and then the increment becomes non-linear. Such a behaviour agrees well with the ChPT predictions~\cite{bib:JB_Kogut_ChPT,bib:DT_Son_ChPT}.

\begin{acknowledgments}
 The authors are grateful to M.~M{\"u}ller-Preussker and V.~G.~Bornyakov for stimulating discussions. Numerical calculations were performed at the ITEP systems "Graphyn" and "Stakan" (authors are much obliged to A.~V.~Barylov, A.~A.~Golubev, V.~A.~Kolosov, I.~E.~Korolko and M.~M.~Sokolov for the help) and at the IHEP cluster (authors are much obliged to V.~V.~Kotlyar for the help). This work was supported by the Russian Foundation for Basic Research (projects nos. 14-02-01185-a, 15-02-07596-a, and 15-32-21117 mol\_a\_ved), by the Council of the President of the Russian Federation for Support of Young Scientists and Leading Scientific Schools (project no. MD-3215.2014.2), and by the FAIR-Russia Research Centre. The work of A.~Yu.~Kotov was supported by the Dynasty Foundation. 
\end{acknowledgments}

\end{document}